\documentclass[epj]{svjour}

\usepackage{graphicx}
\usepackage{fancyhdr}
\def\makeheadbox{{
 }}
\def\mytitle{My title} 
\def\myauthors{My name}  
\def\mytype{My type of session}
\def\mysession{My session}

\def\mytitle{Supersymmetric three dimensional conformal sigma models} 
\def\myauthors{Etsuko Itou}    
\def\mytype{Contributed Talk}    
\def\mysession{Theoretical Models}

\newcommand{\beq}{\begin{eqnarray}}
\newcommand{\eeq}{\end{eqnarray}}
\begin{document}
\title{Supersymmetric three dimensional conformal sigma models}
\subtitle{}
\author{Takeshi Higashi\inst{1}
\thanks{\emph{Email:} higashi@het.phys.sci.osaka-u.ac.jp},%
Kiyoshi Higashijima\inst{1}
\thanks{\emph{Email:} higashij@phys.sci.osaka-u.ac.jp}%
 \and
Etsuko Itou\inst{2}
\thanks{\emph{Email:} itou@yukawa.kyoto-u.ac.jp}%
}                     
%
%
\institute{Department of Physics, 
Graduate School of Science, Osaka University,\\ 
Toyonaka, Osaka 560-0043, Japan \\
\and Yukawa Institute for Theoretical Physics, Kyoto University, \\
Kyoto 606-8502, Japan}
%
\date{Octorber 2007}
\abstract{
We construct supersymmetric conformal sigma models in three dimensions\cite{Higashi:2007tn}.
Nonlinear sigma models in three dimensions are nonrenormalizable in
perturbation theory. We use the Wilsonian renormalization group equation
method, which is one of the nonperturbative methods, to find the fixed points. Existence of fixed points is extremely
important in this approach to show the renormalizability. Conformal sigma
models are defined as the fixed point theories of the Wilsonian
renormalization group equation. The Wilsonian renormalization group
equation with anomalous dimension coincides with the modified Ricci flow
equation. The conformal sigma models are characterized by one parameter which
corresponds to the anomalous dimension of the scalar fields. Any Einstein-K\"{a}hler manifold corresponds to a conformal field theory when the anomalous dimension is $\gamma=-1/2$. Furthermore, we investigate the properties of target spaces in detail for two dimensional case, and find the target space of the fixed point theory becomes compact or noncompact depending on the value of the anomalous
dimension. 
} 
\maketitle
\section{Introduction}
\label{intro}
Nonlinear sigma models with ${\cal N}=2$ supersymmetry in two or three dimensions are defined by the so-called K\"{a}hler potential $K(\phi, \bar{\phi})$, which is a function of the chiral and anti-chiral superfields, $\phi^i$ and $\bar{\phi}^{\bar{j}}$. The bosonic fields $\varphi^i(x)$ play the role of the coordinates of the target manifold ${\cal M}$. 
 The metric, characterizing the target manifold ${\cal M}$, is obtained by the second derivative of this K\"{a}hler potential
\[
g_{i\bar{j}}=\frac{\partial^2K(\varphi, \bar{\varphi})}{\partial \varphi^i\partial \bar{\varphi}^{\bar{j}}}\equiv K,_{i\bar{j}}.
\]
The manifold defined by a K\"{a}hler potential is called the K\"{a}hler manifold.
This metric is an arbitrary function of the scalar fields. 
The Lagrangian of nonlinear sigma model with ${\cal N}=2$ supersymmetry reads
\begin{eqnarray}
{\cal L} = g_{i\bar{j}} \partial_{\mu} \varphi^i \partial^{\mu} \bar{\varphi}^{\bar{j}} + i g_{i\bar{j}} \bar{\psi}^{\bar{j}} (D\llap / \psi)^i + 
\frac{1}{4} R_{i\bar{j}k\bar{l}} \psi^i \psi^k \bar{\psi}^{\bar{j}} \bar{\psi}^{\bar{l}},\nonumber\\ \label{sigma-action}
\end{eqnarray}
where the covariant derivative for the fermion fields is given by
\[
(D_{\mu}\psi)^i = \partial_{\mu} \psi^i + \partial_{\mu} \varphi^j \Gamma^i{}_{j k}\psi^k.
\]

The first term of the Lagrangian (\ref{sigma-action}) shows infinite number of the derivative interactions.
In two dimensions, the scalar fields are dimensionless and the Lagrangian are perturbatively renormalizable.
However, in three dimensions, all these interaction terms are perturbatively nonrenormalizable, since the scalar fields have the canonical dimension $d_{\varphi}=1/2$.

We will investigate these two and three dimensional NL$\sigma$Ms using the nonperturbative renormalization group method.

\section{Two dimensional case}
Any NL$\sigma$M is renormalizable within perturbation theories in two dimensions. 
It is convenient 
to use the Wilsonian renormalization group (WRG) equation for the 
nonperturbative study of field theories with infinitely many 
coupling constants. 
In the paper\cite{HI}, we derived the $\beta$ function for 
$2$-dimensional ${\cal N}=2$ supersymmetric NL$\sigma$M using the 
WRG equation, obtaining  
\beq
\beta (g_{i \bar{j}})=\frac{1}{2\pi} R_{i \bar{j}} +\gamma \Big[\varphi^k g_{i \bar{j},k}+\varphi^{* \bar{k}}g_{i \bar{j}, \bar{k}}+2 g_{i \bar{j}} \Big].
\eeq
The WRG equation describes the variation of the Wilsonian effective action 
when the cutoff scale is changed 
\cite{Wilson Kogut}. 
The first term, proportional to 
the Ricci tensor of the target space, comes from the one-loop diagrams, 
whereas the second term, proportional to the anomalous dimension 
$\gamma$ of fields, comes from the rescaling of fields needed to properly normalize 
the kinetic term. The presence of the anomalous dimension 
reflects the nontrivial continuum limit of the fields.

When the anomalous dimension of the field vanishes, 
scale invariance is realized for NL$\sigma$Ms on Ricci-flat 
K\"{a}hler (Calabi-Yau) manifolds \cite{AFM}. Calabi-Yau metrics 
have been explicitly constructed for some noncompact manifolds 
\cite{HKN}, in the case that the number of isometries is sufficient to reduce 
the Einstein equation to an ordinary differential equation. 

However, when the anomalous dimension of the fields does not 
vanish, the condition of scale invariance is 
quite different.  In the paper{\cite{SU2dim}}, we study novel conformal 
field theories with anomalous dimensions by solving for the condition 
at the fixed point: $\beta =0$. We assume ${\bf U}(N)$ symmetry 
to reduce a set of partial differential equations to an ordinary 
differential equation. The conformal theories obtained have one free 
parameter corresponding to the anomalous dimension of the scalar fields.
The geometry of the target manifolds depends strongly on the sign of 
the anomalous dimensions.

In particularly, the solution of the $\beta(g_{i \bar{j}})=0$ is very simple in the case that the target manifold is of one complex dimension.
The properties of the target manifold of the solutions depend strongly on the sign of the parameter $a=-4 \pi \gamma$, here $\gamma$ is the anomalous dimension of the scalar fields.

When $a>0$, the anomalous dimension is negative.
Because the line element is given by in polar coordinates, with $x=re^{i\phi}$,
	\beq
	ds^2=\frac{1}{1+a r^2} \Big( (dr)^2 +r^2 (d \phi)^2   \Big),\label{line-2}
	\eeq
	the volume and the distance from the origin ($r=0$) to infinity ($r=\infty$) are divergent, while the length of the circumference at infinity is finite. Therefore, the shape of the target manifold is that of a semi-infinite cigar. The volume integral of the scalar curvature is also finite, giving the Euler number is equal to that of a disc.
The theory is known as Witten's Eucleaden black hole solution\cite{Witten}.

\begin{figure}[h]
\begin{center}
\includegraphics[width=5cm]{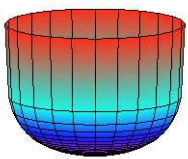}
\caption{The target manifold for $a=+1$ embedded in $3$-dimensional flat Euclidean spaces. It takes the form of a semi-infinite cigar with radius $\sqrt{\frac{1}{a}}$. }
\label{fig:Euclidean}
\end{center}
\end{figure}

	Figure \ref{fig:Euclidean} shows the manifold embedded in $3$-dimensional flat Euclidean spaces.
	The distance between any two points is measured along the shortest path on the surface in the Euclidean space.
	
\begin{figure}[h]
\begin{center}
\includegraphics[width=5cm]{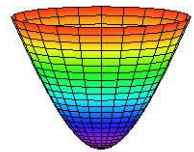}
\caption{The target manifold for $a=-1$, embedded in a flat Minkowski space. The vertical axis has negative signature. In the asymptotic region $\rho \rightarrow \infty$, the surface approaches the lightcone.}
\label{fig:Minkowski}
\end{center}
\end{figure}

When $a<0$, the anomalous dimension is positive. 
	In this case, the metric and scalar curvature read
	\beq
	g_{i \bar{j}}&=&\frac{1}{1-|a|x}, R=\frac{-|a|}{1-|a|x}.
	\eeq
	This metric is ill-defined at the boundary $|x| \sim \frac{1}{\sqrt{-a}}$.
	This is not merely a coordinate singularity, because the scalar curvature is divergent at the boundary.
	Although the volume integral is divergent, the distance to the boundary is finite.
	Now, let us embed this manifold in a flat space.
The manifold is embedded as a space-like surface in a flat Minkowski space.
	Figure \ref{fig:Minkowski} shows the manifold embedded in a $3$-dimensional flat Minkowski space.
	
\section{Renormalizability of three dimensional sigma models}
Renormalization group (RG) equation for the metric of the target manifold ${\cal M}$ in three dimensional sigma models has been derived in \cite{HI,HI3}
\begin{eqnarray}
-\frac{d}{dt} g_{i \bar{j}}
=
\frac{1}{2 \pi^2}R_{i \bar{j}} 
&+&\gamma\left( 2g_{i \bar{j}}+\varphi^k g_{i \bar{j},k} 
+\varphi^{* \bar{k}}g_{i \bar{j},\bar{k}} \right)\nonumber\\
&+&\frac{1}{2}\left(\varphi^k g_{i \bar{j},k} 
+\varphi^{* \bar{k}}g_{i \bar{j},\bar{k}} \right).
\nonumber
\end{eqnarray}
This RG equation, derived by using the so-called K\"{a}hler normal coordinate\cite{KNC}, can be written in a covariant form 
\begin{eqnarray}
-\frac{d}{dt} g_{i \bar{j}}
=\frac{1}{2 \pi^2}R_{i \bar{j}}-g_{i\bar{j}}
+\nabla_i\xi_{\bar{j}}+\nabla_{\bar{j}}\xi_i\label{eq:ricciflow}
\end{eqnarray}
if we define a vector field 
\begin{equation}
\xi^i=\left(\frac{1}{2}+\gamma\right)\varphi^i\label{eq:vector_field}
\end{equation}
in the K\"{a}hler normal coordinate. In other coordinate system, we have to choose a vector field corresponding to the scale transformation of the target manifold.

Let us consider the theories whose target spaces are Einstein-K\"{a}hler manifolds.
The Einstein-K\"{a}hler manifolds satisfy the condition
\beq
R_{i \bar{j}}=\frac{h}{a^2}g_{i \bar{j}},\label{EKcond}
\eeq
where $a$ is the radius of the manifold, which is related to the coupling constant $\lambda$ by $\lambda =\frac{1}{a}$.

Using the Einstein K\"{a}hler condition (\ref{EKcond}), we can obtain the anomalous dimension and $\beta$ function respectively, because only $\lambda$ depends on $t$:
\beq
\gamma&=&- \frac{h \lambda^2}{4\pi^2}.\label{gamma}\\
\beta(\lambda)&\equiv&-\frac{d \lambda}{dt}=-\frac{h}{4\pi^2}\lambda^3+\frac{1}{2} \lambda .\label{beta-lambda}
\eeq
We have an IR fixed point at $\lambda=0$, and we also have a UV fixed point at $\lambda^2=\frac{2 \pi^2}{h}\equiv \lambda_c^2$ for positive $h$. {\it Therefore, if the constant $h$ is positive, it is possible to take the continuum limit by choosing the cutoff dependence of the bare coupling constant as
\beq
\lambda(\Lambda) \stackrel {\Lambda \rightarrow \infty}{\longrightarrow} \lambda_c-\frac{M}{\Lambda},\label{continuum}
\eeq
where $M$ is a finite mass scale.}
With this fine tuning, ${\cal N}=2$ supersymmetric nonlinear $\sigma$ models are renormalizable, at least in our approximation, if the target spaces are Einstein-K\"{a}hler manifolds with positive curvature.

It should be emphasized that although the RG equation obtained in the
perturbation theory has the similar form with the RG equation obtained
in the Wilson's renormalization method, it is valid only in the vicinity
of the free field theory, whereas the Wilsonian RG equation can be used
to study even nontrivial conformal field theories located far away from
the free field theory.

When the constant $h$ is positive, the target manifold is a compact Einstein-K\"{a}hler manifold \cite{Page and Pope}.
In this case, the anomalous dimensions at the fixed points are given by
\beq
\gamma_{IR}&=&0 \mbox{ at the IR fixed point (Gaussian fixed point),}\nonumber\\
\gamma_{UV}&=&-\frac{1}{2} \mbox{at the UV fixed point.}\nonumber
\eeq
At the UV fixed point, the scaling dimension of the scalar fields ($x_{\varphi}$) is equal to the canonical plus anomalous dimensions:
\beq
x_{\varphi}&\equiv& d_{\varphi} + \gamma_{\varphi}=0.
\eeq
Thus the scalar fields and the chiral superfields are dimensionless in the UV conformal theory, as in the case of two dimensional field theories.
Above the fixed point, the scalar fields have non-vanishing mass, and the symmetry is restored.

\begin{figure}[h]
\begin{center}
\begin{picture}(190,100)
\includegraphics[width=5cm]{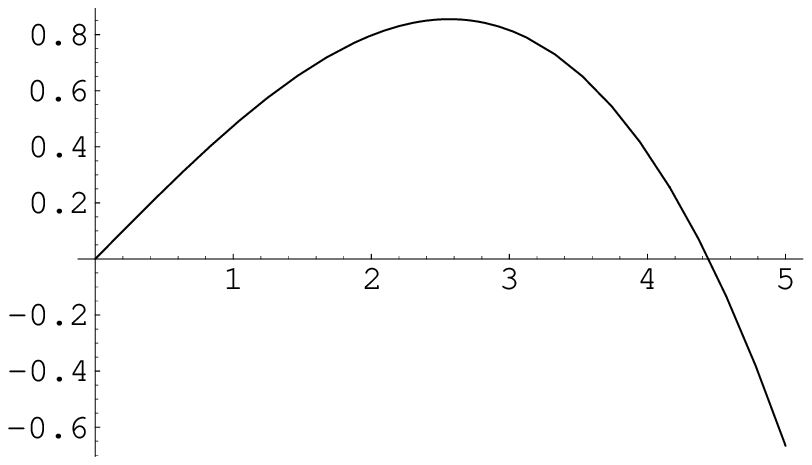}
\put(-150,90){\Large $\beta(\lambda)$}
\put(0,40){\Large $\lambda$}
\end{picture}
\caption{The beta function for Einstein-K\"ahler theories with positive scalar curvature.}
\end{center}
\end{figure}

\section{Conformal sigma models}
The RG equation (\ref{eq:ricciflow}), called the modified Ricci flow in mathematical literature\cite{Chow-Knopf}, describes the deformation of the target manifold of the effective theory.

The fixed point, invariant under the change of the mass scale, is obtained by solving an equation
\begin{equation}
\frac{1}{2 \pi^2}R_{i \bar{j}}-g_{i\bar{j}}
+\nabla_i\xi_{\bar{j}}+\nabla_{\bar{j}}\xi_i=0.\label{eq:cft}
\end{equation}
The metric $g_{i\bar{j}}$ satisfying this equation defines a conformal field theory, and such solution is called the K{\"a}hler-Ricci soliton \cite{Koiso90}.

From now, we put the parameter $c$ as
\beq
c\equiv \frac{1}{2}+\gamma,
\eeq
 which corresponds to the conformal dimension of the scalar fields at fixed point.

Although it is difficult to solve eq.(\ref{eq:cft}) explicitly for $\gamma\ne -\frac{1}{2}$, it can be solved for two-dimensional target space ${\cal M}$ by using a graphical method. We use real variables to describe the target manifold ${\cal M}$, and choose a special gauge where the line element of ${\cal M}$ takes the following form
\begin{equation}
ds^2=dr^2+e^2(r)d\phi^2.\label{eq:synchronousgauge}
\end{equation}
Since our target spaces are complex manifolds, we have assumed rotational symmetry in the $\phi$ direction corresponding to the $U(1)$ symmetry, and normalize the range of $\phi$ to $0\le \phi <2\pi$. Then $e(r)$ denotes the radius of a circle for a fixed value of $r$.

The fixed point of the RG equation written in terms of real coordinates corresponds to the solution of
\begin{equation}
a^2R_{ij}-g_{ij}
+\nabla_i\xi_j+\nabla_j\xi_i=0,\label{eq:cft_real}
\end{equation}
where 
\[
a^2=\frac{1}{2\pi^2}.\label{eq:radius_sphere}
\]
Now, we have to find the vector field $\xi^i=(\xi^r, \xi^{\phi})$. 
The vector field $\xi^i$, representing an infinitesimal scale transformation of the target space, has to be proportional to $(cr,0)$ at least around the origin $r=0$, that is renorlization condition to normalize the kinetic term in the RG equation. Since we assume the rotational symmetry ($U(1)$), it is natural to assume $\xi_{\phi}=0$. Then the vector field in this coordinate system is fixed by the consistency of the coupled differential equation (\ref{eq:cft_real}).
We obtain the RG equation in this gauge
\begin{equation}
-a^2e''-e + 2cee'=0.\label{eq:rg_synchronous}
\end{equation}

When $c=0$, namely for $\gamma=-1/2$, the solution of this equation is easily obtained
\[
e(r)=a\sin{\frac{r}{a}},
\]
which defines the line element of the round $S^2$ with radius $a$. 

On the other hand, when $c\ne 0$, it is convenient to rewrite the second order 
differential equation to a set of the first order differential equations
\begin{eqnarray}
e'&=&p \label{eq:1st_order}\\
p'&=&-\frac{1}{a^2}e(1-2cp)\nonumber
\end{eqnarray}
with the boundary condition
\begin{equation}
e(0)=0,\quad p(0)=1\label{eq:bc_sg}
\end{equation}

\begin{figure}[h]
\begin{center}
\unitlength=1mm
\begin{picture}(60,30)
\includegraphics[width=5cm]{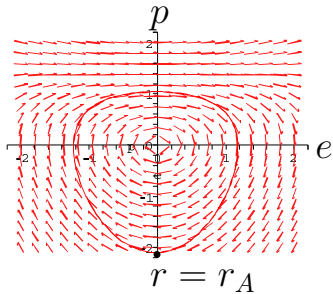}
\put(-26,35){\Large $p$}
\put(-4,17){\Large $e$}
\put(-26,0){\Large $r=r_A$}
\put(-26.3,2){\Huge $\cdot$}
\end{picture}
\caption{Flow of the first order differential equations (\ref{eq:1st_order}) for $0< 2c<1$ in the``phase space" $(e(r),p(r))$. 
The solid line represents the solution specified by the boundary condition.}\label{fig:dsphere}

\end{center}
\end{figure}
\begin{figure}[h]
\begin{center}
\unitlength=1mm
\begin{picture}(60,40)
\includegraphics[width=5cm]{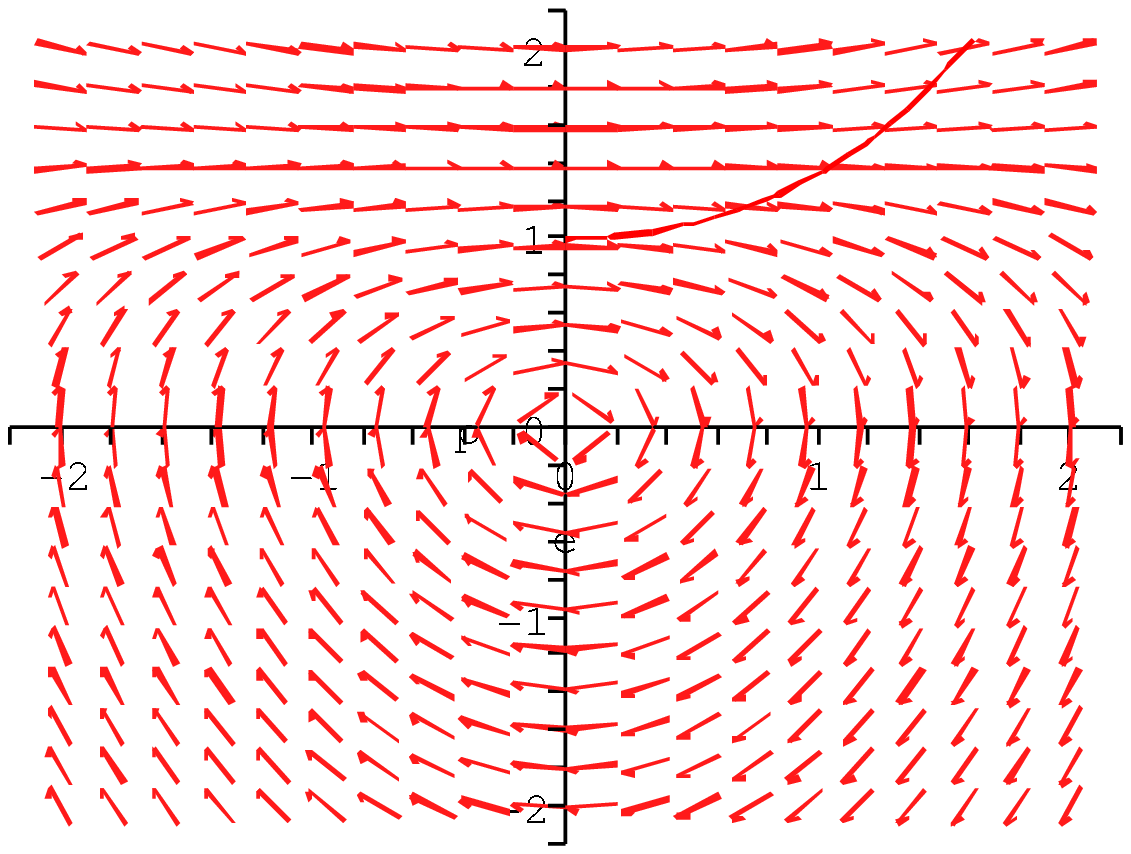}
\put(-26,35){\Large $p$}
\put(-4,17){\Large $e$}
\end{picture}
\caption{Flow of the first order differential equations (\ref{eq:1st_order}) for $2c\geq1$ in the phase space. 
The solid line represents the solution specified by the boundary condition.}\label{fig:non-cpt}

\end{center}
\end{figure}
The vector field of the flow (\ref{eq:1st_order}) is shown in Fig.\ref{fig:dsphere}. When $0\le 2c < 1$, this equation defines a compact manifold since the trajectory starting from the initial point (\ref{eq:bc_sg}) comes back to $e=0$ at a finite $r=r_A$ implying the circumference of the circle at that $r$ vanishes. 
We may call it the ``deformed shpere". Similarly, we have compact target space for $c<0$.

On the other hand, Fig.\ref{fig:non-cpt} shows the radius $e(r)$ becomes larger and larger when $r$ goes to infinity, then the solution corresponds to a noncompact manifold 
for $2c\ge 1$. 
To see the asymptotic behavior of $e(r)$ for large $r$, we will be able to neglect the second term in (\ref{eq:rg_synchronous}) 
\[
\frac{de}{dr}=1+\frac{c}{a^2}e^2,
\]
which can be integrated to obtain $e(r)$ 
\[
e(r)=\frac{a}{\sqrt{c}}\tan{\left(\frac{\sqrt{c}}{a}r\right)}.
\]
Since $e(r)$ defines the radius of the circle when the geodesic distance from the origin $r$ is fixed, this asymptotic behavior implies that the manifold for $2c>1$ has an increasingly large radius for large $r$.

%
%

\section*{Acknowledgment}
The author (E.I.) is supported by a Grant-in-Aid for the 21st Century COE ``Center for Diversity and Universality in Physics". 
This work was supported in part by Grants-in-Aid for Scientific Research
(\#340075).

%

\begin{thebibliography}{999}
%
%


\bibitem{Higashi:2007tn}
  T.~Higashi, K.~Higashijima and E.~Itou,
Prog.\ Theor.\ Phys.\  {\bf 117} (2007) 1139  
[arXiv:hep-th/0702188].


\bibitem{HI} K.~Higashijima and E.~Itou, 
Prog.\ Theor.\ Phys.\  {\bf 108} (2002) 737, {\tt  hep-th/0205036}.


\bibitem{Wilson Kogut} K.G.~Wilson and I.G.~Kogut,
Phys. Rep. {\bf 12} (1974) 75.
F.~Wegner and A.~Houghton,
Phys. Rev. {\bf A8} (1973) 401.
T.R.~Morris,
Int.\ J.\ Mod.\ Phys.\ A {\bf 9} (1994) 2411, {\tt hep-ph/9308265};
K.~Aoki,
Int.\ J.\ Mod.\ Phys.\  {\bf B14} (2000) 1249.


\bibitem{AFM} L.~Alvarez-Gaum\'{e},D.Z.~Freedman and S.~Mukhi,
Ann.of Phys. $\bf{134}$ (1981) 85.

\bibitem{HKN} K.~Higashijima, T.~Kimura and M.~Nitta,
Nucl. Phys. {\bf B645} (2002) 438, {\tt hep-th/0202064}.

\bibitem{SU2dim} K.~Higashijima and E.~Itou,
Prog.\ Theor.\ Phys.\  {\bf 109} 751, {\tt hep-th/0302090}.


\bibitem{Witten} E.~Witten, Phys. Rev. {\bf D44} (1991) 314.

\bibitem{HI3}
  K.~Higashijima and E.~Itou,
  Prog.\ Theor.\ Phys.\  {\bf 110} (2003) 563
  [arXiv:hep-th/0304194].


\bibitem{KNC} K.~Higashijima and M.~Nitta, Prog.\ Theor.\ Phys.\ {\bf 105} (2001) 243,
{\tt hep-th/0006027}.\\
 K.~Higashijima, E.~Itou and M.~Nitta,
Prog.\ Theor.\ Phys.\  {\bf 108} (2002) 185,
 {\tt hep-th/0203081}.

\bibitem{Page and Pope} D. N.~Page and C.N.~Pope, 
Class. Quant. Grav. {\bf 3} (1986) 249.


\bibitem{Chow-Knopf}
B.~Chow and D.~Knopf,
The Ricci Flow: An Introduction, American Mathematical Society, 2004.
I.~Bakas, Comptes Rendus Physique {\bf 6} (2005) 175. 

\bibitem{Koiso90}
N.~Koiso,
Recent topics in differential and analytic geometry, 327-337, Adv.~Stud.~Pure~Math., 18-I, Academic Press, Boston, MA, 1990.\\
H-G.~Cao,
J. Differential Geom. {\bf 45} (1997) no.2, 257.\\
T.~Ivey,
Proc. Amer. Math. Soc. {\bf 125} (1997) no.4, 1203.









\end{thebibliography}
%

\end{document}